\documentclass[aps,prb,twocolumn,superscriptaddress]{revtex4}
\usepackage[usenames]{color}
\newcommand{\comment}[1]{}
\usepackage{graphicx}
\usepackage{amsmath}
\usepackage{epsfig}
\usepackage{rotating}

\def\be{\begin{eqnarray}}
\def\ee{\end{eqnarray}}

\def\nairo{Na$_4$Ir$_3$O$_8$}

\begin{document}

\bibliographystyle{revtex}
\title{Spin-orbit coupling in the metallic and spin-liquid phases of Na$_4$Ir$_3$O$_8$}

\date{\today}

\author{Daniel Podolsky}
\affiliation{Department of Physics, University of Toronto,
Toronto, Ontario M5S 1A7, Canada}
\affiliation{Department of Physics, Technion, Haifa 32000, Israel}

\author{Yong Baek Kim}
\affiliation{Department of Physics, University of Toronto,
Toronto, Ontario M5S 1A7, Canada}

\begin{abstract}
It has recently been proposed that Na$_4$Ir$_3$O$_8$ is a weak Mott insulator at ambient pressure, supporting a three-dimensional spin liquid phase with a spinon Fermi surface. This proposal is consistent with recent experimental findings  that the material becomes  a metal upon increasing pressure or doping. In this work, we investigate the effect of the spin-orbit coupling arising from 5$d$ Ir moments both in the metallic and spin liquid phases of  Na$_4$Ir$_3$O$_8$. The effective Hubbard model in terms of pseudospin $j=1/2$ Ir states is derived and its consequences to both metallic and spin liquid phases are studied. In particular, the model leads to enhanced Wilson ratio and strong temperature dependence of the Hall coefficient. 
\end{abstract}

\maketitle

\section{Introduction}

Recently, \nairo\ has emerged as a promising candidate for a three dimensional quantum spin-liquid. Susceptibility measurements show a large antiferromagnetic Curie-Weiss temperature in this material, $\Theta_{CW}=-650$ K, yet no evidence of magnetic ordering has been observed down to very low temperatures \cite{okamoto}.  The material at ambient pressure is an insulator, yet specific heat measurements support the existence of a finite density of low energy excitations.  This has lead to different theoretical proposals for Mott insulator states with fermionic spinon excitations \cite{lawler08a,lawler08b,palee08,PPKS}, and to proposals to detect the spinon Fermi surface \cite{NormanMicklitz}.

At a superficial level, a three dimensional material may be an unlikely candidate for a quantum spin-liquid.  Fluctuations are usually weak in three dimensions, and magnetic order is likely to set in at finite temperatures.  However, despite its three-dimensional nature, \nairo\ combines a confluence of factors which favour the formation of a spin liquid: the Ir atoms have spin-1/2, leading to strong quantum fluctuations; these atoms are arranged in an undistorted ``hyperkagome'' lattice, which gives rise to magnetic frustration; and, perhaps most importantly, there are strong charge fluctuations in \nairo.  These charge fluctuations lead to ring-exchange-type interactions of increasingly long-range between moments, and can prevent the formation of long-range order altogether\cite{motrunich05}.

The strongest evidence for charge fluctuations in \nairo\ comes from recent measurements on samples under pressure and on doped samples\cite{takagi08}.  Application of hydrostatic pressure enhances the conductivity of the undoped samples by several orders of magnitude, indicating the proximity of \nairo\ to a metallic phase.  This is further corroborated by the fact that relatively small concentrations of dopants turn the system metallic \cite{takagi08}.  This indicates that \nairo\ is a weak Mott insulator, which can readily undergo bandwidth and doping-controlled transitions to the nearby metallic phase.

The weak Mott insulator scenario for \nairo\ was first proposed in Ref.~\onlinecite{PPKS}, in which the bandwidth-tuned transition to a metal was studied in detail.  In this theory, the electrons are factorized into fermionic spinons, which carry spin but no charge, and bosonic rotors, which carry charge but no spin \cite{florens04}. When a critical interaction strength is reached, the rotors become gapped, and the system becomes an insulator.  On the other hand, the spinons remain gapless even in the insulator, forming a neutral spinon Fermi sea, responsible for the finite density of excitations seen in the specific heat.
The spin-liquid states proposed so far, however, have ignored the sizeable spin-orbit coupling effects present in \nairo\cite{chen08}.  The heavy Ir atoms have a large spin-orbit coupling which must be included in the same footing as electronic hopping and correlation effects in any microscopic theory of \nairo.  

In this paper, we study the effects of spin-orbit coupling on the weak Mott insulator and metallic phases of \nairo.  We find that the system is well-described by a one-band Hubbard model, in which the electrons occupy {\em total angular momentum} $j=1/2$ (``pseudospin'') states on sites of the hyperkagome  lattice.  Thus, the situation is similar to the absence of spin-orbit coupling, but with two new ingredients: instead of carrying spin $s=1/2$, the electrons carry pseudospin $j=1/2$, and the hopping between sites does not necessarily conserve pseudospin.  However, the fact that the system is still described by a one-band model is significant, as it justifies the general approach followed in Ref.~\onlinecite{PPKS}.

%The extended 5$d$ Ir orbitals give rise to a large crystal field splitting between $e_g$ and $t_{2g}$ orbitals.  For Ir$^{4+}$, the $e_g$ orbitals remain empty, whereas the five outer-shell $d$ electrons leave a single vacancy in the $t_{2g}$ levels. The $t_{2g}$ levels are further split by spin-orbit coupling into a completely filled $j=3/2$ multiplet and a half-filled $j=1/2$ multiplet.  This pseudospin $j=1/2$ multiplet forms the band

The importance of spin-orbit in \nairo\ can be appreciated from recent studies on other transition metal oxides containing iridium \cite{kim08,jin08,moon08,kim09,shitade09,jin09,pesin09}.  For example, one may expect that correlation effects in Sr$_{n+1}$Ir$_n$O$_{3n+1}$ would be weak due to the extended nature of the 5d Ir orbitals. Instead, spin orbit coupling gives rise to a narrow $j=1/2$ band in these materials\cite{moon08}.  In Sr$_2$IrO$_4$ this leads to an antiferromagnetic Mott state, as demonstrated through angle resolved photoemission spectroscopy, optical conductivity, and x-ray absorption spectroscopy measurements, and also through first-principles electronic structure calculations \cite{kim08,jin08}.  We find that a similar mechanism is responsible for enhancing correlation effects in \nairo.

The inclusion of spin-orbit coupling may also help explain one of the outstanding experimental observations in \nairo.  At low temperatures in the Mott insulator phase, the specific heat has a linear temperature dependence, with a finite coefficient $\gamma\equiv C/T$, as in a metal.  Similarly, the magnetic susceptibility is constant at low temperatures, as in a metal.  These results may be evidence for the existence of a spinon Fermi surface in the insulator\cite{PPKS}.  On the other hand, in the experiments the Wilson ratio between the susceptibility and the specific heat coefficient is much greater than one.  This is often taken to be a signature of strong ferromagnetic fluctuations, but it seems to be at odds with the antiferromagnetic Curie-Weiss temperature \cite{okamoto}.  The inclusion of spin-orbit coupling gives an alternate way of generating a Wilson ratio different from one\cite{chen08}, which we will explore below.

Our approach in this paper is partially phenomenological.  After deriving an effective Hubbard model based on microscopic considerations, we are left with a single free parameter $\theta$ that controls the ratio of direct Ir-Ir hopping to the hopping mediated by oxygens.  Since the outer 5d electrons in Ir are extended and it is difficult {\em a priori} to make an estimate of $\theta$, instead we follow a phenomenological approach and estimate $\theta$ by requiring a large Wilson ratio and a small $\gamma$.  However, many of our qualitative results are independent of the precise value $\theta$.  For instance, we find that for all values of $\theta$, the non-interacting system has a finite density of states.  This is in contrast to the pyrochlore lattice, where spin-orbit effects can lead to the formation of a topological band insulator\cite{pesin09}.  Therefore, the insulating behaviour in \nairo\ must arise due to interaction effects.

Another generic feature that is independent of the detailed value of $\theta$ is a strong temperature dependence in many physical quantities.  The hyperkagome  lattice has a 12 site unit cell which, together with the presence of spin-orbit coupling, leads to 24 quasiparticle bands.  We find that the Fermi surface contains multiple pockets arising from many of the bands, which have very strong structure near the chemical potential.  This leads to a strong dependence of the density of states with energy, which is reflected in the temperature dependence, {\em e.g.} of the specific heat and also of the Hall coefficient for doped samples.

This paper is organized as follows.  In Section \ref{sec:Hubbard} we give a microscopic derivation of the one band Hubbard model 
of \nairo in terms of the pseudospin $j=1/2$ Ir states.  In Section III, we analyze the quasiparticle spectrum in the non-interacting limit
and study the resulting thermodynamics and transport properties in the metallic phase. We discuss the implications of
the spin-orbit coupling in the spin liquid insulator. 

\section{Hubbard model}
\label{sec:Hubbard}

Our starting point in deriving a microscopic model for \nairo\ is the observation that due to the large atomic number of Ir, the spin-orbit coupling $\lambda_{so}$ at the Ir sites is expected to be comparable to other microscopic energy scales, such as the local Hubbard repulsion $U$ and the hopping amplitude between neighbouring Ir sites, and therefore must be treated on the same footing.  Therefore, electronic spin $S$ is not expected to be a good quantum number.  In what follows, we first consider a single Ir site with spin-orbit coupling, which we will then use to construct a tight binding model.

The Ir 5d orbitals are split into $e_g$ and $t_{2g}$ levels by the octahedral crystal field of the nearby oxygens.  The $e_g-t_{2g}$ splitting is large relative to $\lambda_{so}$ \cite{chen08,kim08,jin08}.  Therefore, the $e_g$ orbitals in Ir$^{4+}$ are completely empty and can be ignored, whereas the $t_{2g}$ orbitals have a single $s=1/2$ hole.  The $t_{2g}$ orbitals are weakly split due to weak distortions of the oxygen octahedra.  However, we assume that this splitting is much smaller than $\lambda_{so}$. Then, the three $t_{2g}$ orbitals behave as an effective orbital angular momentum $L=1$ moment.  The spin-orbit coupling between this $L=1$ moment and the $s=1/2$ spin yields a low energy $j=3/2$ multiplet and a high energy $j=1/2$ multiplet, separated by the energy $\lambda_{so}$.  Provided that the electronic hopping between Ir atoms is not large relative to $\lambda_{so}$, the $j=3/2$ and $j=1/2$ bands remain well-separated.  Then, the $j=3/2$ band is completely full and the $j=1/2$ band is the only one to participate in the low energy physics.  The emergence of a $j=1/2$ band in an iridium oxide has been studied in detail in the closely related material Sr$_2$IrO$_4$ \cite{kim08,jin08,kim09}

Hence, $j=1/2$ states give a good description of Ir electrons in \nairo.  The $j=1/2$ states are represented in the $j_z=\pm 1/2$ basis by,\cite{chen08}
\begin{eqnarray}
|\uparrow_j\rangle &=& \frac{1}{\sqrt{3}}\left(i|xz,\downarrow_s\rangle+|yz,\downarrow_s\rangle+|xy,\uparrow_s\rangle) \right. , \nonumber\\
|\downarrow_j\rangle &=& -\frac{1}{\sqrt{3}}\left(i|xz,\uparrow_s\rangle-|yz,\uparrow_s\rangle+|xy,\downarrow_s\rangle) \right. , \nonumber
\end{eqnarray}
where we have used the subscripts $j$ and $s$ to distinguish between total and spin angular momenta labels.  
By comparison, in the absence of spin-orbit coupling, the description would be in terms of $s=1/2$ moments.  In what follows, we need to keep track of whether the labels $|\uparrow\rangle$ and $|\downarrow\rangle$ represent spin angular momentum $s$, or ``pseudospin'' $j$.  We will use the labels $\alpha=\pm 1$ and $\sigma=\pm 1$ to describe pseudospin and spin, respectively, {\em  i.e.} $|\alpha=+1\rangle=|\uparrow_j\rangle$, $|\alpha=-1\rangle=|\downarrow_j\rangle$, $|\sigma=+1\rangle=|\uparrow_s\rangle$, and $|\sigma=-1\rangle=|\downarrow_s\rangle$.  We can now introduce interactions and hopping, to obtain a one-band Hubbard model on the hyperkagome  lattice:
\begin{eqnarray}
H=H_0+ \frac{U}{2}\sum_i n_i^2 - \mu\sum_{i} n_i \ .
\label{eq:Hubbard}
\end{eqnarray}
Here, $d_{i\alpha}$ annihilates a $d$ electron on the Ir site $i$, $n_i=\sum_\alpha d_{i\alpha}^\dagger d_{i\alpha}$ is the total density of electrons occupying the $j=1/2$ multiplet on that site, and $H_0$ is a non-interacting tight-binding Hamiltonian,
\begin{eqnarray}
H_0=-\sum_{ij\alpha\alpha'} t_{i\alpha,j\alpha'}d_{i\alpha}^\dagger d_{j\alpha'} \ .
\label{eq:tightbinding}
\end{eqnarray}
The pseudospin-dependent hopping amplitude $t_{i\alpha,i\alpha'}$ is derived below.

\subsection{Tight-binding parameters}

\begin{figure}
\includegraphics[width=3.2in]{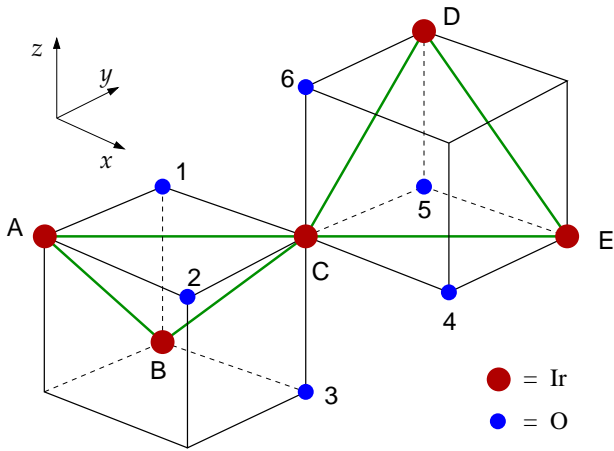}
\caption{(Color online) Neighbourhood of one Ir site (C) in the hyperkagome  compound \nairo. Site C is surrounded by an octahedron of oxygens (labelled $1\ldots6$), and it has four nearest neighbour Ir sites, A, B, D, and E. We ignore the weak cubic distortions present in \nairo.  Note that site A has only one second neighbour shown in the figure (D) and one third neighbour (E), whereas site B has two second neighbours shown (D and E), and no third neighbours. 
\label{fig:oct} } \vskip-0.15in
\end{figure}

Once we consider many sites, the hopping amplitudes become pseudospin-dependent due to spin-orbit coupling.  To derive the effective tight-binding Hamiltonian for the $j=1/2$ bands, we assume three types of hopping processes
are present: {\em (i)} Direct hopping between $d$-orbitals on the nearest neighbour Ir sites ({\em e.g.} sites A and C in Fig.~\ref{fig:oct}:
\begin{eqnarray}
H_{AC}&=&-t_1 d_{A,xy,\sigma}^\dagger d_{C,xy,\sigma}+t_2 d_{A,xz,\sigma}^\dagger d_{C,yz,\sigma}\nonumber\\
&\,& +t_2d_{A,yz,\sigma}^\dagger d_{C,xz,\sigma}+h.c. 
\end{eqnarray}
{\em (ii)} Hopping between Ir $d$-orbitals and the nearest neighbour O $p$-orbitals ({\em e.g.} between sites A and 1):
\begin{eqnarray}
H_{A1}=-t_{dp} d_{A,xy,\sigma}^\dagger p_{1x\sigma}-t_{dp} d_{A,yz,\sigma}^\dagger p_{1z\sigma}+h.c. 
\end{eqnarray}
and {\em (iii)} Hopping between nearest neighbour O $p$-orbitals ({\em e.g.} between sites 1 and 5):
\begin{eqnarray}
H_{15}&=&-t_\pi p_{1z\sigma}^\dagger p_{5z\sigma}-\frac{t_\pi-t_\sigma}{2}\left(p_{1y\sigma}^\dagger p_{5y\sigma}+p_{1x\sigma}^\dagger p_{5x\sigma}\right)\nonumber\\
&\,&-\frac{t_\pi+t_\sigma}{2}\left(p_{1y\sigma}^\dagger p_{5x\sigma}+p_{1y\sigma}^\dagger p_{5x\sigma}\right)+h.c. %\\
%	H_{16}&=&-t_\pi p_{1x\sigma}^\dagger p_{6x\sigma}-\frac{t_\pi-t_\sigma}{2}\left(p_{1y\sigma}^\dagger p_{6y\sigma}+p_{1z}^\dagger p_{6z\sigma}\right)\nonumber\\
%&\,&-\frac{t_\pi+t_\sigma}{2}\left(p_{1y\sigma}^\dagger p_{6z\sigma}+p_{1z\sigma}^\dagger p_{6y\sigma}\right)+h.c.\\
%H_{26}&=&-t_\pi p_{2y\sigma}^\dagger p_{6y\sigma}-\frac{t_\pi-t_\sigma}{2}\left(p_{2x\sigma}^\dagger p_{6x\sigma}+p_{2z\sigma}^\dagger p_{6z\sigma}\right)\nonumber\\
%&\,&+\frac{t_\pi+t_\sigma}{2}\left(p_{2x\sigma}^\dagger p_{6z\sigma}+p_{2z\sigma}^\dagger p_{6x\sigma}\right)+h.c.
\end{eqnarray}
In these expressions, summation over repeated spin indices $\sigma$ is implicit. Note that we have defined 5 independent hopping parameters, $t_1$, $t_2$, $t_{dp}$, $t_{\pi}$, and $t_{\sigma}$, where the last two describe $\pi$ and $\sigma$ hopping processes
between nearest-neighbour $p$-orbitals.

Most of the hopping processes relevant to \nairo\ involve virtual states with vacancies in the oxygen $p$-orbitals.  These processes are suppressed by  the energy denominator $E_p \equiv \epsilon_d-\epsilon_p-U_p >0$, where $\epsilon_d$ and $\epsilon_p$ are the level energies of the $j=1/2$ iridium orbitals and the oxygen $p$ orbitals, respectively, and $U_p$ is the Hubbard repulsion on the oxygens.  The energy $E_p$ is assumed to be large relative to the hopping amplitude between iridium and oxygen, and also to the hopping between two oxygens.  Therefore, the energy $E_p$ controls the perturbation theory used to compute the effective tight binding model parameters, which are expressed in powers of $t/E_p$.

Symmetry arguments provide strong constraints on the form of the tight-binding model that obtains from perturbation theory.  For instance,
since spin-orbit coupling is time-reversal invariant, the most general pseudospin-dependent hopping appearing in Eq.~(\ref{eq:tightbinding}) is of the form
\begin{eqnarray}
t_{i\alpha,j\alpha'}=t^0_{ij}\delta_{\alpha\alpha'}+i{\bf v}_{ij}\cdot {\vec{\sigma}}_{\alpha\alpha'} , 
\end{eqnarray}
where ${\bf v}_{ij}$ is a real vector and ${\vec{\sigma}}=(\sigma_x,\sigma_y,\sigma_z)$ are the three pseudospin Pauli $\sigma$ matrices.  Note that ${\bf v}_{ij}$ must be odd under exchange of $i$ and $j$ in order to ensure Hermiticity of $H_0$.  In addition, note that ${\bf v}$ transforms as a vector under rotations within the point symmetry group of the lattice.  

Consider parity transformations about some point ${\bf r}_0$, defined by ${\cal P}_{{\bf r}_0}:{\bf r}_0+{\bf r}\to {\bf r}_0-{\bf r}$. Since pseudospin is parity invariant ({\em i.e.} angular momentum is a pseudovector), the effect of a parity transformation ${\cal P}_{(i+j)/2}$ with origin at the midpoint between sites $i$ and $j$ is
\begin{eqnarray}
{\cal P}_{(i+j)/2}: {\bf v}_{ij}\to {\bf v}_{ji}=-{\bf v}_{ij} \ .
\end{eqnarray}
Therefore, if the lattice is invariant under ${\cal P}_{(i+j)/2}$, then ${\cal P}_{(i+j)/2}^\dagger H_0 {\cal P}_{(i+j)/2}=H_0$, which implies that ${\bf v}_{ij}=0$, {\em i.e.} the hopping between sites $i$ and $j$ is pseudospin symmetric.  In practice, the hyperkagome  lattice has an intrinsic handedness, and therefore is, in general, not invariant under any parity transformation.  However, at low orders in $t/E_p$, many of the hopping processes involve parity-invariant clusters of sites, leading to pseudospin symmetric hopping amplitudes.

\subsubsection{Nearest neighbor hopping}

To leading order, the nearest neighbor hopping between sites $A$ and $C$ involves the creation of an electron at site A with pseudospin $\alpha$, hopping to site C through the term $H_{AC}$ in the Hamiltonian, and projection into a final pseudospin $\alpha'$ state:
\begin{eqnarray}
t_{C\alpha',A\alpha}&=&-\langle C\alpha'|H_{AC}|A \alpha\rangle\nonumber\\ &=&-\frac{1}{3}\left(-it_2+it_2-t_1\right)\delta_{\alpha\alpha'}\nonumber\\
&=&\frac{1}{3}t_1\delta_{\alpha\alpha'} \ .
\label{eq:nearest1}
\end{eqnarray} 
Note that, at this level in perturbation theory, the nearest neighbor hopping is isotropic in pseudospin.  This can be understood from the symmetry considerations discussed above -- clearly, the cluster composed of the two sites A and C is parity invariant about its midpoint.  Note that nearest-neighbor hopping mediated by oxygens vanishes to second order in perturbation theory.  Thus, the leading correction to Eq.~(\ref{eq:nearest1}) occurs at third order in perturbation theory, and is given by
\begin{eqnarray}
t_{C\alpha',A\alpha}=\frac{1}{3}\left(t_1+ \frac{t_{dp}^2(t_\sigma+3t_\pi)}{E_p^2} \right)\delta_{\alpha\alpha'} \ .
\label{eq:nearest2}
\end{eqnarray}
Even in this case, the perturbation theory processes involve parity invariant clusters, and therefore only represent a weak pseudospin-symmetric renormalization of the result in Eq.~(\ref{eq:nearest1}).  The leading pseudospin-dependent terms involve next-nearest neighbor hopping, which we study next.
 
\subsubsection{Second neighbor hopping}

To third order in perturbation theory, hopping between sites A and D can occur through three independent paths:
\begin{eqnarray}
A\to 1\to 6\to D , \nonumber\\
A\to 2\to 6\to D , \nonumber\\
A\to 1\to 5\to D . \nonumber
\end{eqnarray}
Among these, the first path actually involves a parity symmetric cluster.  However, the other two do not, and therefore give rise to pseudospin-dependent hopping.  Summing the contributions from all three paths yields,
\begin{eqnarray}
t_{D\alpha',A\alpha}=\frac{t_{dp}^2(t_\sigma-t_\pi)}{3 E_p^2}\delta_{\alpha\alpha'}-i{\bf v}_{DA}\cdot{\vec{\sigma}}_{\alpha'\alpha} ,
\label{eq:nextnearest}
\end{eqnarray}
where
\begin{eqnarray}
{\bf v}_{DA}=\frac{t_{pd}^2}{3E_p^2} \left(\begin{array}{c} t_\pi-t_\sigma \\ 2(t_\pi+t_\sigma)  \\ t_\pi-t_\sigma \end{array}\right) \ .
\end{eqnarray}
The hopping amplitude from $E$ to $B$ can be obtained from Eq.~(\ref{eq:nextnearest}) through a $C_2$ rotation at site $C$, which yields 
\begin{eqnarray}
{\bf v}_{BE}=\frac{t_{pd}^2}{3E_p^2} \left(\begin{array}{c}  -2(t_\pi+t_\sigma) \\ -(t_\pi-t_\sigma) \\ -(t_\pi-t_\sigma) \end{array}\right) , 
\end{eqnarray}
and for hopping from $B$ to $D$, by doing an inversion about $C$, followed by a proper rotation (this operation is a symmetry for the clusters involved at this order in perturbation theory).  This leads to,
\begin{eqnarray}
{\bf v}_{DB}=\frac{t_{pd}^2}{3E_p^2} \left(\begin{array}{c} -(t_\pi-t_\sigma) \\ -(t_\pi-t_\sigma)  \\ -2(t_\pi+t_\sigma) \end{array}\right) \ .
\end{eqnarray}
 
\subsubsection{Third neighbor hopping}

The third neighbor hopping between sites $A$ and $E$ involves 16 different paths.  The calculation is significantly simplified by the observation that to the lowest order in perturbation theory (third order), third neighbor hopping involves a reflection-symmetric cluster, and is therefore pseudospin symmetric.  We find,
\begin{eqnarray}
t_{E\alpha',A\alpha}=\frac{t_{pd}^2(t_\pi+t_\sigma)}{3 E_p^2} \ . \label{eq:thirdneighbor}
\end{eqnarray}

\subsection{Parametrization of the hopping amplitudes}

Inspection of the equations (\ref{eq:nearest2}), (\ref{eq:nextnearest}), and (\ref{eq:thirdneighbor}) shows that the hopping amplitudes can be parametrized as,
\begin{eqnarray}
t_1 &=& t \cos\theta , \\
t_\pi \frac{t_{pd}^2}{E_p^2} &=& t \sin\theta , \\
t_\pi&=&\eta |t_\sigma|.
\end{eqnarray}
where $t$ represents an overall energy scale for the hopping Hamiltonian, $\eta$ is the dimensionless ratio between $t_\pi$ and $|t_\sigma|$, and $\theta$ is an angle representing the relative strength of direct Ir-Ir hopping and hopping mediated by oxygens.  We expect $\eta$ to be a small parameter.  In our computations, we find that its precise value is not very important, and take $\eta=0.1$ for concreteness.  

On the other hand, it is difficult to make an {\em a priori} estimation of the angle $\theta$.  In the case of Ir, the outer 5d electrons are extended and there can be in principle a sizeable orbital overlap between adjacent Ir atoms.  Hence, we do not try to estimate the microscopic value of the parameter $\theta$, which controls the relative size of direct Ir-Ir hopping and hopping mediated by the oxygens.  Instead, in this analysis we will take $\theta$ to be a phenomenological parameter, which will be obtained from a comparison with experimental data.  However, as we will emphasize, many of the qualitative results of our analysis do not depend on the detailed value of $\theta$.

\section{Non-interacting limit}
\label{sec:thermodynamic}

In the previous section, we derived a Hubbard model for \nairo, see Eq.~(\ref{eq:Hubbard}).  In this section, we will compute various physical observables of this model in the non-interacting limit.  One reason to focus on this limit is that experimental evidence indicates that the system becomes metallic under hydrostatic pressure and by doping\cite{takagi08}.  Therefore, many of the properties of the metallic state may be captured by the non-interacting model, up to quasiparticle renormalization effects.  Moreover, the non-interacting limit can also give us relevant information for the insulating state.  As was argued in Ref.~\onlinecite{PPKS}, the insulating state is a weak Mott insulator, described by gapped rotors coexisting with a gapless Fermi surface of spinons.  In this picture, the spinons would have a similar dispersion to the electronic dispersion obtained in the non-interacting model.  Later on, we will discuss some of the effects of reintroducing interactions.

\subsection{Spectrum}

\begin{figure}
\includegraphics[width=3.4in]{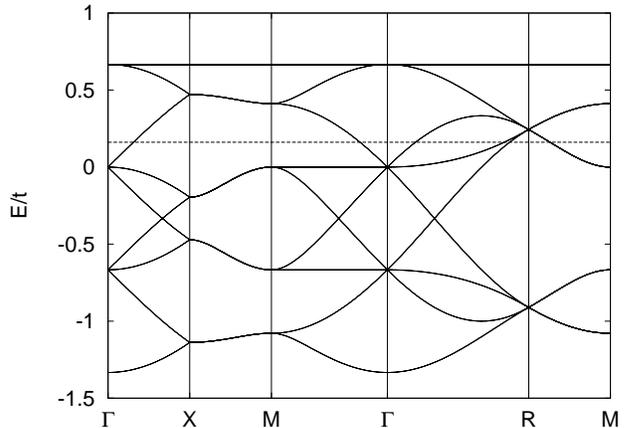}
\caption{Spectrum along high-symmetry directions, for the case with nearest-neighbor hopping only $\theta=0$.  The momentum labels are $\Gamma=(0,0,0)$, $X=(\pi,0,0)$, $M=(\pi,\pi,0)$, and $R=(\pi,\pi,\pi)$.  There are eight degenerate flat bands at  the top of the spectrum, $\epsilon=2 t$. The chemical potential at half-filling is shown as a dashed line.
\label{fig:t0s} } \vskip-0.15in
\end{figure}

\begin{figure}
\includegraphics[width=3.4in]{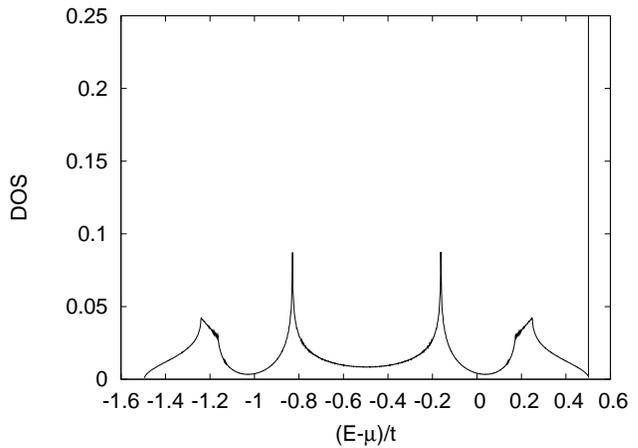}
\caption{Density of states for a system with nearest-neighbor hopping only $\theta=0$.  The energies are shifted by the chemical potential at half-filling.  The delta function at high end of the spectrum arises from the flat bands.
\label{fig:t0d} } \vskip-0.15in
\end{figure}

The unit cell in the hyperkagome  lattice consists of 12 sites.  Hence, when taking into account the pseudospin degree of freedom, we obtain 24 different quasiparticle bands in the Brillouin zone.  Figure \ref{fig:t0s} shows the quasiparticle spectrum for the case $\theta=0$, {\em i.e.} with nearest-neighbour hopping only.  In this case, the system is pseudospin symmetric, so that pseudospin up and down bands are always degenerate.  At the high end of the spectrum, there are eight degenerate flat bands (four for each pseudospin), corresponding to localized states that arise due to the geometric frustration of the lattice.  At half-filling, for each value of pseudospin, the chemical potential cuts through one small electron-like pocket and two small hole-like pockets.
Figure \ref{fig:t0d} shows the associated density of states.  Note that near the chemical potential the density of states is small.  This is due to the small size of the electron and hole pockets at the chemical potential, and leads to a strong temperature dependence of the specific heat in the nearest-neighbour model seen in Refs.~\onlinecite{lawler08b} and \onlinecite{palee08}.

\begin{figure}
\includegraphics[width=3.4in]{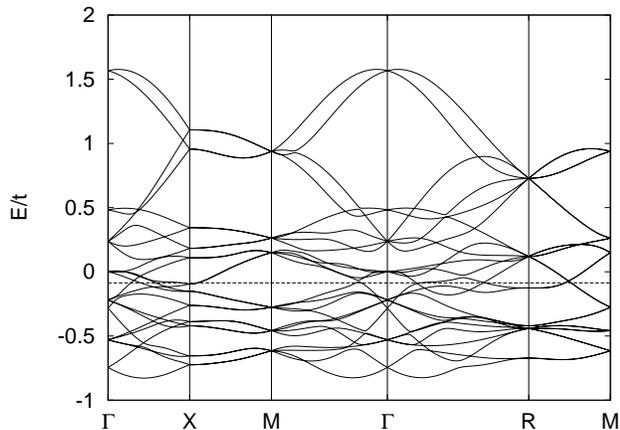} 
\caption{Spectrum for $\theta=3.31$. Note that many of the degeneracies in Fig.~\ref{fig:t0s} have been lifted by furthest neighbour hopping. The chemical potential at half-filling is shown as a dashed line.
\label{fig:t2s} } \vskip-0.15in
\end{figure}

\begin{figure}
\includegraphics[width=3.4in]{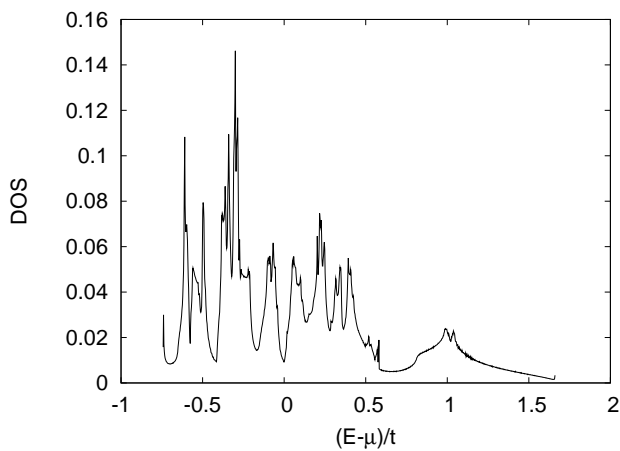}
\caption{Density of states for $\theta=3.31$.  Energies measured relative to the chemical potential at half-filling
\label{fig:t2d} } \vskip-0.15in
\end{figure}

For $\theta\ne 0$ the hopping becomes pseudospin-dependent. Figures \ref{fig:t2s} and \ref{fig:t2d} show the quasiparticle spectrum and associated density of states for the choice of parameter $\theta=3.31$.  Note that many of the degeneracies in Fig.~\ref{fig:t0s} have been lifted, although some residual degeneracy is left at the high symmetry points.  We have verified that the degeneracies at the $\Gamma=(0,0,0)$ point are consistent with the symmetry of the magnetic point group of the hyperkagome  lattice.  Also note that, due to the further neighbour hopping, the flat bands seen in Fig.~\ref{fig:t0s} have now acquired a finite dispersion.    

%Note that for this value of $\theta$, a group of degenerate states at point $R=(\pi,\pi,\pi)$ lie very close to the chemical potential at half-filling. The only sizeable pockets in this case surround the $\Gamma$ point.  Hence, the density of states at the Fermi level in Fig.~\ref{fig:t2d} is very small.

At half-filling, there is a sizeable electron pocket near the $X=(\pi,0,0)$ point, and a large number of smaller pockets with shallow energy dispersions.  These include a small electron pocket near $(\pi,0,0)$, as well as hole pockets near $(0,0,0)$, and $(\pi/2,\pi,\pi)$.  The existence of multiple shallow pockets leads to a strong energy dependence of the density of states  and to a suppressed density of states at the chemical potential, as shown in Fig.~\ref{fig:t2d}.  This energy dependence is reflected in many of the physical quantities computed below.   Although we only present results for two different values of $\theta$, we find that the strong energy dependence in the density of states seen in Fig.~\ref{fig:t0d} for $\theta=0$, and in Fig.~\ref{fig:t2d} for $\theta=3.31$, is generically present for most values of $\theta$.  This arises due to the large unit cell {of the hyperkagome lattice}, which leads to multiple narrow bands that straddle the chemical potential.

\subsection{Thermodynamics}

 There are two related experimental facts in \nairo\ which we will take as guides to set the parameter $\theta$: first, the specific heat coefficient $\gamma\equiv C/T$, is very small; and second, the Wilson ratio $W$, defined by $W\equiv \frac{\pi^2}{3}\frac{\chi/\mu_B^2}{\gamma/k_B^2}$, is much greater than one.  Here, $\chi$ is the low temperature paramagnetic susceptibility.

\begin{figure}
\includegraphics[width=3.4in]{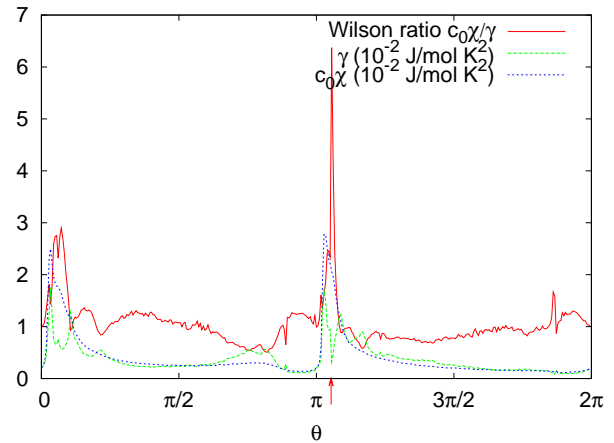}
\caption{Low temperature thermodynamic quantities as a function of the parameter $\theta$.  Shown are the Wilson ratio $\frac{\pi^2}{3}\frac{\chi/\mu_B^2}{\gamma/k_B^2}$, the specific heat coefficient $\gamma=C/T$, and the susceptibility (scaled by $c_0\equiv\pi^2 k_B^2/3\mu_B^2$).  The value $\theta=3.31$ is indicated by a vertical arrow. For concreteness, we choose the overall hopping amplitude to be $t=200$ meV.
\label{fig:theta} } \vskip-0.15in
\end{figure}

Figure \ref{fig:theta} displays the Pauli spin susceptibility $\chi$, the specific heat coefficient $\gamma$,  and the Wilson ratio $W$, at low temperatures as a function of $\theta$.  For $\theta=0$, the specific heat coefficient is very small, but the Wilson ratio is equal to one.  This is expected, as the normalization of the Wilson ratio is chosen such that an isotropic non-interacting Fermi gas has $W=1$.  On the other hand, as $\theta$ is increased, the Wilson ratio becomes greater than one.  However, this occurs at the price of an increased $\gamma$.  At the value $\theta=3.31$ (indicated by an arrow in Fig.~\ref{fig:theta}), we obtain a good compromise between a large Wilson ratio and a small $\gamma$.  This value of $\theta$ is in qualitative agreement with the electronic structure predicted by density functional theory calculations\cite{NormanLDA}. In what follows, we will focus on this value of $\theta.$

\begin{figure} 
\includegraphics[angle=270,width=3.4in]{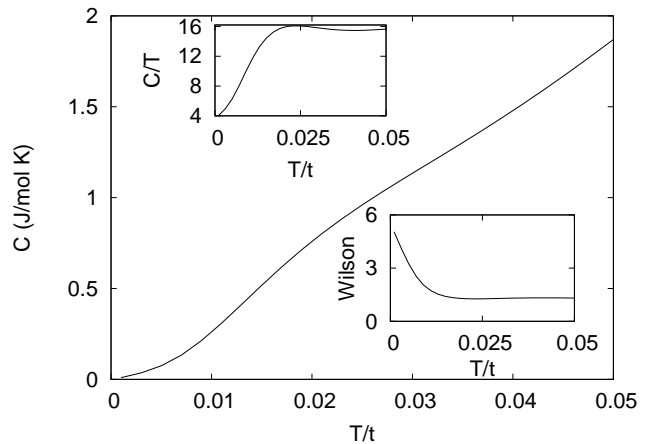}
\caption{Specific heat versus temperature at $\theta=3.31$.  Insets: Wilson ratio and $C/T$ versus $T$ for $\theta=3.31$.  $C/T$ in the inset is measured in units of mJ/mol K$^2$, for $t=200$ meV.
\label{fig:cv} } \vskip-0.15in
\end{figure}

Figure \ref{fig:cv} shows the temperature dependence of the specific heat at $\theta=3.31$.  Note that $C$ has significant curvature in the low temperature regime, which arises due to the strong energy dependence of the density of states shown in Fig.~\ref{fig:t2d}.  The top inset in Fig.~\ref{fig:cv} shows the ratio $C/T$ as a function of temperature, which displays a strong suppression at low temperatures.   The bottom inset shows the Wilson ratio, which increases strongly as temperature is reduced.  This temperature dependence is dominated by the specific heat, as the susceptibility is approximately constant over this temperature range.

%Note the small value of the zero temperature specific heat coefficient $\gamma=C/T$, due to the small density of states at the Fermi level.  On the other hand, the density of states grows quickly as we move away from the Fermi level, giving rise to a strong $T^2$ contribution to the specific heat.  As $t_\sigma$ is increased, the coefficient of the $T^2$ contribution increases steadily, while $\gamma$ remains roughly constant.  It is only at the largest value of $t_\sigma$ considered, $t_\sigma=0.3t$, that $\gamma$ increases markedly.

%Figure \ref{fig:wr} compares the zero temperature value of $\gamma$ to the Pauli spin susceptibility, and also shows the Wilson ratio of these two quantities, scaled by a factor $\pi^2/3$.  Note that the Wilson ratio displays a strong enhancement near $t_\sigma=0.22$.

\subsection{Transport in doped samples}

Recent experiments indicate that doping \nairo\ renders the system metallic\cite{takagi08}.  To investigate the transport properties in such a metal, we consider the electrical and Hall conductivities in dopings ranging from 1\% to 4\%, which is controlled by varying the chemical potential.  We then compute transport properties using the Boltzmann equation results\cite{OngHall}:
\begin{eqnarray} 
\sigma_{xx}&=&e^2\tau \frac{1}{V}\sum_{a,\bf k} \left(-\frac{\partial f}{\partial \epsilon^a_{\bf k}}\right)(v^a_x)^2 , \label{eq:boltzmann}\\
\sigma_{xy}&=&e^3 \tau^2 H_z \frac{1}{V}\sum_{a,\bf k}\left(\frac{\partial f}{\partial \epsilon^a_{\bf k}}\right)v^a_y({\bf v^a}\times {\bf \nabla_{\bf k}})_{\hat{z}} v^a_x , \nonumber
\end{eqnarray}
where $a=1\ldots 24$ is a band index and ${\bf v}^a_{\bf k}=\nabla_{\bf k}\epsilon$ is the fermion velocity of electrons in band $a$. Here we have assumed that the scattering time $\tau$ is independent of the wave vector ${\bf k}$ and also of the band index $a$.

\begin{figure}
\includegraphics[width=3.4in]{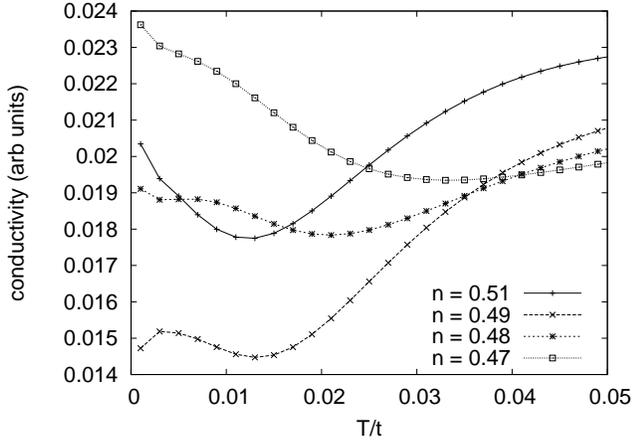}
\caption{Conductivity for doping away from half-filling for $\theta=3.31$.  The conductivity is shown in arbitrary units -- by Eq.~(\ref{eq:boltzmann}), $\sigma_{xx}$ is proportional to a scattering time $\tau$ which is not known.
\label{fig:sxx} } \vskip-0.15in
\end{figure} 

Fig.~\ref{fig:sxx} shows electrical conductivity as a function of temperature, for various doping levels. Notice that for small doping away from half-filling, the electrical conductivity is a monotonically {\em increasing} function of temperature.  This is an artifact of the assumption that the scattering time $\tau$ used in Eq.~(\ref{eq:boltzmann}) is a constant.  In other words, in these calculations we are only taking into account the elastic scattering contribution to $\tau$, and ignoring the temperature dependent inelastic scattering processes.  Thus, the temperature dependence in Fig.~\ref{fig:sxx} is a density of states effect.  More realistically, in real samples at high temperatures, the electrical conductivity will decrease with heating.  However, it is possible that at lower temperatures the electrical conductivity may display this density-of-states-induced conductivity increase.  This effect should be more pronounced for dirty samples, for which the elastic scattering dominates over the inelastic scattering over a broader range of temperatures.

\begin{figure}
\includegraphics[width=3.4in]{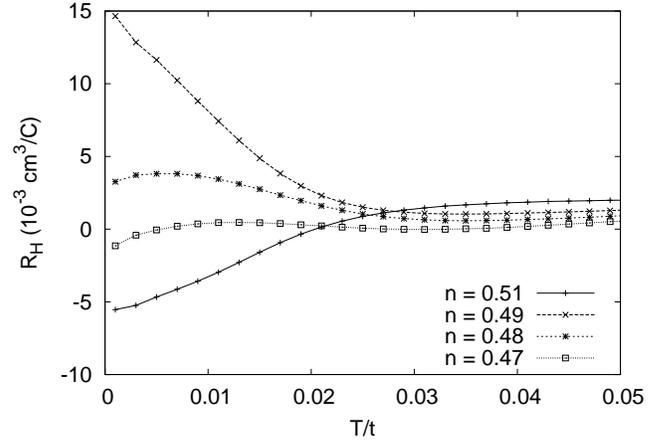}
\caption{Hall coefficient for doping away from half-filling for $\theta=3.31$
\label{fig:rh} } \vskip-0.15in
\end{figure}

Figure \ref{fig:rh} shows the Hall coefficient $R_H$ as a function of temperature for $\theta=3.31$.  In contrast to the electrical conductivity, the Hall coefficient $R_H\approx-\frac{\sigma_{xy}}{\sigma_{xx}^2 H_z}$ is independent of scattering time, at least for wave vector-independent $\tau$.  Note that there is a change from electron-like to hole-like $R_H$ that is mediated by doping.  For all dopings, we find that the Hall coefficient is a strong function of temperature.  As with other observables studied before, this effect is due to the presence of multiple narrow bands near the chemical potential.  Thus, although the detailed form of $R_H$ does depend on $\theta$ and also on doping, the fact that it is a strong function of temperature is a generic feature of 
the model in Eq.(\ref{eq:Hubbard}).

\section{Discussion on the Spin Liquid Phase}

Although the tight-binding spectrum is derived for electrons in a metal, the same tight binding model would describe the spinon spectrum in the
spin liquid phase up to renormalization effects coming from the interaction between spinons and the gauge field.
It was shown that this interaction only leads to weak renormalization effects near the critical point between the metal
and the spin liquid\cite{PPKS}. For example, it only leads to a $\ln \ln 1/T$ enhancement of the specific heat coefficient $C/T$.\cite{PPKS}
On the other hand, the previous analysis did not include the spin-orbit coupling and it is expected that 
there would be an additional spin-orbit-coupling-induced renormalization of specific heat and susceptibility. 
That is, just like the metallic phase discussed earlier,
the Wilson ratio would be in general bigger than one: the same arguments with the pseudospin-dependent
hopping amplitudes apply to the spinons in the spin liquid phase. 

There is, however, even further renormalization of the Wilson ratio due to the combined effect of the spin-orbit coupling 
and the quasiparticle (electrons or spinons) interactions. In the framework of the Landau Fermi liquid theory,
it was shown that additional quasiparticle interactions that depend on both the orbital and spin quantum
numbers can arise in spin-orbital coupled systems. In particular, an additional Fermi liquid renormalization factor for
the Wilson ratio has the form: {$(1+G_1) \over (1+F_0)(1+G_1)-(G_2)^2$}, where $F_0$ corresponds to the
usual spin density-spin density interaction, and $G_1$ and $G_2$ represent additional spin-orbit coupling-induced 
quasiparticle interactions that vanish in the absence of the spin-orbit coupling\cite{Quader}. 
Given that the nature of the underlying quasiparticle interactions among
spinons is quite different from that between electrons, it is conceivable that the Wilson ratio in metal and
the spin liquid can also be quantitatively quite different. Since the interactions between spinons are generally 
believed to be much stronger, we may expect a bigger Wilson ratio in the spin liquid phase.
Quantitative estimation would require the derivation of the full Fermi liquid interaction function, which
may be an excellent topic of future study.

%Thus Eq.~(\ref{eq:boltzmann}) can also be used to compute the spinon contribution to the conductivity.  
%By the Ioffe-Larkin rule, the spinon conductivity must be added to the rotor conductivity in parallel, {\em i.e.} the resistivities of the 
%two channels must be added.  The rotor conductivity near the metal-insulator transition was discussed in Ref.~\ref{PPKS}.  There we found that in the insulator, the resistivity of the system is activated, whereas in the quantum critial region there is a $1/T$ divergence in resistivity, {\em i.e.} the critical point itself is a weak insulator.

%-Log corrections to $C$ and $\chi$?

\acknowledgements 

We thank R. S. Perry and H. Takagi for showing us unpublished data on Na$_4$Ir$_3$O$_8$ and many inspiring discussions.
We are also grateful to Arun Paramekanti, Nic Shannon, Leon Balents, and Mike Norman for sharing their insight with us.
This work was supported by the NSERC of Canada, the Canada Research Chair program,
and the Canadian Institute for Advanced Research. We acknowledge warm hospitality at the University of Tokyo,
the RIKEN, the Kavli Institute for Theoretical Physics, and
the Aspen Center for Physics, where various parts of this work were performed.

\end{document}